\documentclass[prl,aps,epsf,twocolumn,showpacs]{revtex4-1}
\usepackage[pdftex]{graphicx}
\usepackage{dcolumn}
\usepackage{bm}
\usepackage{epsfig}
\usepackage{latexsym}
\usepackage{amsmath}
\usepackage{amssymb}
\usepackage{color}
\usepackage{array}

\newcommand{\<}{\langle}
\newcommand{\e}{\varepsilon}

\renewcommand{\>}{\rangle}
\renewcommand{\(}{\left(}
\renewcommand{\)}{\right)}
\renewcommand{\[}{\left[}
\renewcommand{\]}{\right]}
\renewcommand{\v}[1]{\mathbf{#1}} 

\newcommand{\bs}{\boldsymbol}

\begin{document}
\title{Quantum Oscillations from Surface Fermi-Arcs in Weyl and Dirac Semi-Metals}
\author{Andrew C. Potter, Itamar Kimchi, and Ashvin Vishwanath}
\affiliation{Department of Physics, University of California, Berkeley, CA 94720, USA}

\begin{abstract}
The surface states of Weyl semi-metals (SM's) consist of disjointed Fermi-arcs.  This unusual surface-Fermiology provides a fingerprint of the topological features of the bulk Weyl-phase.   Using a combination of semiclassical and numerical methods, we show that, in contrast to naive expectation, there are closed magnetic orbits involving the open- surface Fermi-arcs.  Below a critical field strength that depends on sample thickness, these orbits produce periodic quantum oscillations of the density of states in a magnetic field, enabling a variety of experimental probes of the unconventional Fermi-arc surface states.  The orbits are also essential for reproducing the bulk chiral anomaly in a finite slab.  These results are then extended to the closely related and recently discovered 3D Dirac SM materials, including Cd$_3$As$_2$ and Na$_3$Bi, which are doubled copies of Weyl semi-metals protected by crystal symmetry.  Despite the fact that the protecting crystal symmetry is broken by a surface, we show that Dirac materials can still host unconventional surface-states, which can be detected in quantum oscillations experiments.
\end{abstract}
\maketitle

Weyl semi-metals (SMs) are three-dimensional materials for which the bulk band-gap closes at an even-number of discrete points (Weyl nodes) in the Brillouin zone\cite{WanTurner,TurnerReview}.  Near the Weyl nodes, electrons have relativistic dispersion $\e_k \approx  \pm v|\v{k}|$.  Each Weyl node acts as a monopole or anti-monopole of Berry-curvature, (i.e. a closed surface in momentum space surrounding a node will be pierced by $\pm 2\pi$ Berry flux) and is associated with a positive or negative chirality respectively.  Consequently, 2D cross sections of the Brillouin zone change Chern number by $\pm 1$ across each Weyl node, implying that a generic surface will exhibit surface states whose ``Fermi-surfaces" consist of a set of open line-segments\cite{WanTurner,TurnerReview,DuncanPlumbing}.  

These Fermi-arcs connect pairs of bulk Weyl nodes with opposite chiralities and cannot be removed without annihilating the bulk Weyl nodes. Such unusual surface-Fermiology would be impossible in a purely two-dimensional system, whose Fermi-surface is necessariliy smooth, and cannot abrubtly terminate at a point within the Brillouin zone. Consequently, the Fermi-arcs serve as a surface fingerprint of the topological character of the bulk band-structure, and it is interesting to ask how they might be experimentally observed. 

Traditionally, the most powerful methods of mapping out a material's Fermi-surface rely on periodic-in-$\frac{1}{B}$ quantum oscillations of the density of states in a magnetic field, $B$.  Such quantum oscillations require closed magnetic orbits for electrons at the Fermi-surface, which naively cannot arise from disjointed Fermi-arcs.  This raises the interesting question: do Fermi-arcs lead to quantum oscillations? 

\begin{figure}[ttt]
\begin{center}
\includegraphics[width = 2.7in]{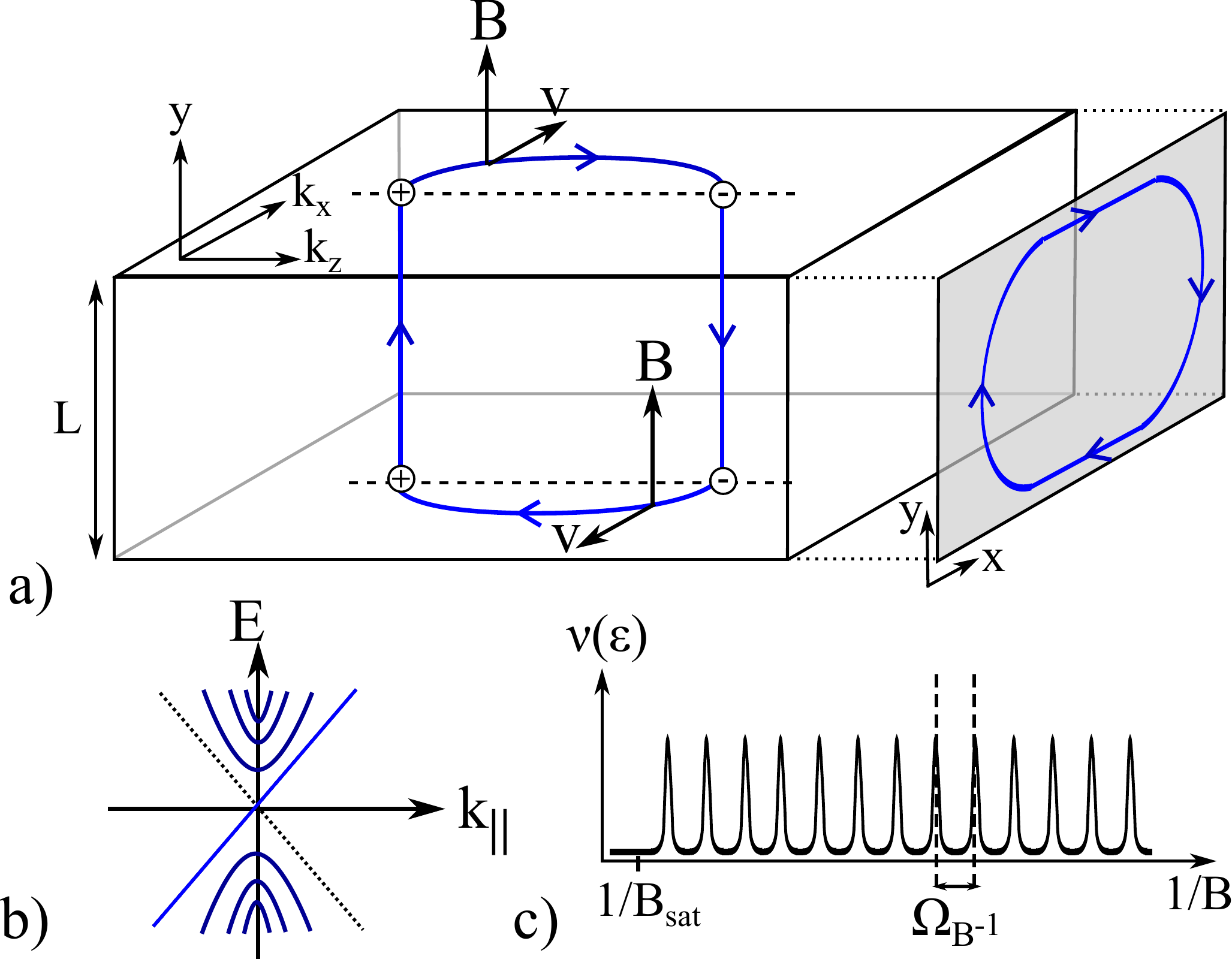}
\end{center}
\vspace{-.2in}
\caption{(a) Semiclassical orbit in a magnetic field along $\hat{y}$, involving surface states that gives rise to quantum oscillations in a finite thickness slab (shown in mixed real space in $y$ and momentum space in $x,z$ directions; inset shows corresponding real-space trajectory); (b) bulk Landau level (LL) spectrum for a `+'-chirality Weyl node, $k_\parallel$ denotes momentum along the direction of the field; (c) periodic-in-$1/B$ features in the density of states resulting from quantizing the orbits shown in (a). The train of peaks ends at a saturation field B$_{\rm sat}$ which scales inversely with the sample thickness $L$. 
}
\vspace{-.2in}
\label{fig:SemiclassicalTrajectory}
\end{figure} 

We answer this question in the affirmative by noting that, in a slab of Weyl semi-metal of finite-thickness, closed magnetic orbits can be obtained by traversing the Fermi-arc on the top surface and returning along the corresponding arc on the bottom surface. Using semiclassical methods, we find that such closed orbits can indeed lead to periodic-in-$\frac{1}{B}$ quantum oscillations of the density of states.  Since the orbits involved require an electron to traverse the bulk in order to connect top and bottom surfaces, the oscillations only occur up to a critical field strength which depends on slab thickness.  For larger fields, the orbits no longer produce periodic quantum oscillations, but are important for understanding the fate of the bulk chiral Landau-levels (LLs) and associated chiral anomaly in finite systems.  The semiclassical results are then validated by direct numerical simulation of the spectrum of a Weyl SM slab in a field.

Weyl SMs are theoretically predicted to occur in strongly spin-orbit coupled systems like the iridates, certain spinels and topological insulator-ferromagnet multilayers\cite{WanTurner,WitczakKrempa,WitczakKrempaReview,ChenHermele}, Yet, despite promising indications\cite{Julian}, there is currently no clearcut experimental candidate. However, two recently discovered materials Cd$_3$As$_2$\cite{WangCd3As2,Cava,Hasan} and Na$_3$Bi\cite{WangNa3Bi,NaBiArpes,CavaHasan} are thought to be 3D Dirac SMs, close cousins of Weyl SMs.  The bulk band-structure of these Dirac SMs consists of two superimposed copies of Weyl
SMs with opposite chiralities. Ordinarily, when Weyl
nodes of opposite chirality are not separated in momentum, they can mix and gap each other out. However, in Dirac SMs intermixing is symmetry forbidden, since the superimposed Weyl nodes belong to different representations of the crystal symmetry (namely, discrete screw-symmetries for Cd$_3$As$_2$ and Na$_3$Bi\cite{WangCd3As2,WangNa3Bi,NaBiArpes,CavaHasan}). 

\begin{figure}[ttt]
\begin{center}
\includegraphics[width = 3.5in]{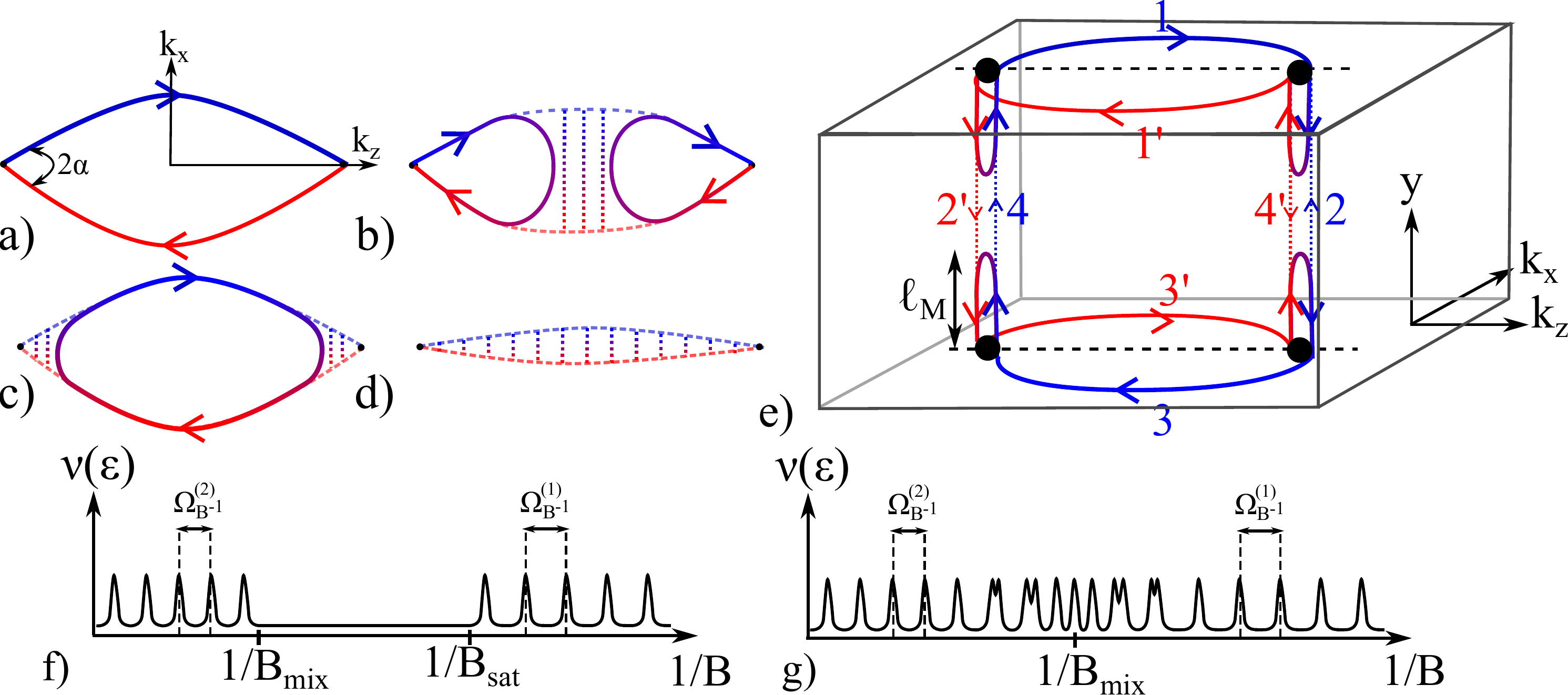}
\end{center}
\vspace{-.2in}
\caption{Viewing the Dirac SM as two superposed copies of a Weyl SM with opposite chiralities suggests two sets of surface arcs, labeled R (blue) and R' (red), that curve in opposite directions and meet at the bulk Weyl nodes with a discontinuous kink (a). Crystal symmetry breaking perturbations mix R and R' (vertical dashed lines) and depending on their strength may either preserve- (a) reconstruct- (b,c) or remove- (d) the surface states.  (e) Semiclassical orbits for Dirac semi-metal with kinked surface states of type (a) (a similar picture holds for those shown in (b)) as two copies of the Weyl SM orbits that are mixed in the bulk do to the
magnetic field.  (f,g) Schematic representation of kinked surface-state (shown in a,b) contributions to the density of states in a field for $B_\text{sat}<B_\text{mix}$ (f) and $B_\text{sat}>B_\text{mix}$ (g).
}
\vspace{-.2in}
\label{fig:DiracCase}
\end{figure}

We show that the known Dirac SM materials can have pairs of surface arcs that meet at a sharp corner or ``kink" at the bulk Dirac nodes (Fig.~\ref{fig:DiracCase}a).  Such a kink would not be allowed in a purely 2D metal, and is a special feature of the crystal-symmetry-protected Weyl structure of the Dirac SM.   This result is not a priori obvious, since the surfaces of interest do not preserve the symmetry that protects the bulk Dirac nodes.  Despite this, we show that surface states and kink feature are perturbatively stable to the symmetry breaking potential of the surface, and can exist so long as this potential is not too strong.  Applying the semiclassical results for the Weyl SM, we describe the signatures of these unconventional Dirac surface states in quantum oscillations experiments.
%

\vspace{4pt}\noindent{\bf Semiclassical Analysis - }
To start, let us consider the simplest case of a single pair of bulk Weyl nodes with chiralities $\pm 1$ located at $\v{k} = \pm K_W\hat{z}$ and a magnetic field $B$ along the $y$-direction (the extension to more complicated cases is straightforward).  In an infinite system, the magnetic field produces Landau-level (LL) bands that disperse only along the field direction.  For $k_W\ell_B\gg 1$, where $\ell_B = \frac{1}{\sqrt{eB}}$ is the magnetic length, the $\pm$-chirality Weyl nodes are effectively decoupled, and the LL spectrum includes gapped, non-chiral LL bands with energies $\e^\pm_n(k_y) \approx \pm \text{sgn}(n)v\sqrt{2\ell_B^{-2}+k_y^2}$ with $n=\pm 1, \pm 2, \dots$ and gapless chiral modes with energies $\e^\pm_0(k_y) = \pm vk_y$, for the $\pm$-chirality nodes respectively.
  
Now consider a slab of Weyl semi-metal that is infinite in the x- and z- directions and with finite-thickness, $L$, along y.  Further, suppose that the slab is sufficiently thick to neglect direct tunneling between states in the center of the Fermi-arcs on the top and bottom surfaces.  Initially we focus at low energies near the bulk Weyl node where the relevant states are those of the surface-Fermi arc and the bulk chiral LLs.  Later, we will see that the results extend to energies well above the Weyl node where the bulk consists of two Fermi-pockets connected at the surface by Fermi-arcs\cite{DuncanPlumbing}.

In a semiclassical description, an electron at z-momentum $k_z$ along the Fermi-arc of the top surface slides along the arc towards the `-' chirality Weyl node according to:
\begin{align} \partial_t \v{k} = -e\v{v}_{\v{k}}\times \v{B} = e v B\hat{t}_{\v{k}} \label{eq:SemiclassicalEOM}\end{align}
Here, we have taken the velocity, $v$, on the Fermi-arc to be  independent of $\v{k}$, and $\hat{t}_\v{k}$ is the unit tangent vector to the arc (with orientation on top and bottom surfaces indicated by the arrows in Fig.~~\ref{fig:SemiclassicalTrajectory}a).  As an electron sliding along the top-surface arc nears the `-' bulk Weyl node,  the energy gap to bulk bands vanishes.  This necessarily leads to breakdown in the single-band semiclassical description in which the electron is transferred from the surface arc into the bulk.  At low energies, the only available bulk states are those of the gapless bulk chiral LL \cite{WanTurner} propagating towards the bottom surface\cite{Endnote:SurfaceBulkTransition}.  This chiral mode acts as a one-way ``conveyor-belt", transporting the electron from the Fermi-arc on the top surface to that of the bottom surface.  Upon reaching the bottom-surface Fermi-arc, the electron then slides to the `+' Weyl node, where it connects with the upwards moving chiral bulk LL, thereby returning to the top surface and completing the orbit (see Fig.~\ref{fig:SemiclassicalTrajectory}c).


%

Quantum energy levels can be approximately obtained from semiclassical orbits that satisfy the condition:
$\e_n t \approx 2\pi (n+\gamma)$ with $n\in \mathbb{Z}$, where $t$ is the semiclassical time associated with the orbit, and $\gamma$ is a constant of order unity encoding low-$n$ quantum effects.  Sliding along the top and bottom Fermi-arcs takes time: $t_\text{arc}\approx \frac{k_0}{evB}$ where  $k_0$ is the arc-length of the Fermi-arc. Propagation between top and bottom surfaces via the bulk chiral LL's takes time $t_\text{bulk}\approx \frac{L}{v}$.  Combining these expressions gives:
\begin{align} \e_n = \frac{ \pi v(n+\gamma)}{L+k_0\ell_B^2} \label{eq:SemiClassicalQuantization}\end{align}
where zero-energy corresponds to the Weyl nodes.

These magnetic orbits involving Fermi-arcs are distinguished from conventional magnetic orbits in ordinary 2D systems or in surface states of 3D systems by the dependence on the slab thickness, $L$.  As explained below, this characteristic $L$ dependence can be extracted by analyzing the dependence of quantum oscillations on field direction. Another peculiar and unconventional feature is that the real-space trajectory of these orbits need not enclose any area perpendicular to the magnetic field (as occurs when the Fermi-arcs are straight lines).

\vspace{4pt}\noindent{\bf Quantum Oscillations - }  Having established the existence of quantized magnetic orbits involving Fermi-arcs we now turn to the question of whether these orbits produce quantum oscillations.  Suppose we fix chemical potential $\mu$, and vary magnetic field.   Eq.~\ref{eq:SemiClassicalQuantization} dictates that the $n^\text{th}$ energy level crosses $\mu$ when:
\begin{align} \frac{1}{B_n} = ek_0^{-1}\(\frac{\pi 
v}{\mu}(n+\gamma)-L\) \label{eq:Binv} \end{align}
where the solution is defined only for sufficiently large $n$ such that the right-hand side is positive.  States cross $\mu$ at regularly spaced intervals in $\frac{1}{B}$ of size $\Omega_{B^{-1}} \approx \frac{e\pi v}{\mu k_0}$.  Each time a level passes through $\mu$, a peak occurs in the density of states, giving rise to periodic-in-$\frac{1}{B}$ oscillations in many measurable quantities like conductivity and magnetization.  These oscillations are analogous to those of an ordinary 2D metal with quadratic dispersion with effective mass $m_\text{eff} = \frac{k_0}{\pi v}$ (though, of course, the bulk Weyl electrons are massless and $m_\text{eff}$ is just an effective parameter with dimensions of mass).  The periodic train of peaks persists only up to fields of order $B_\text{sat} = \frac{k_0}{L}\(\frac{\lceil \frac{\mu L}{\pi v}-\gamma \rceil}{\frac{\mu L}{\pi v}-\gamma }-1\)^{-1}\approx \frac{k_0}{L}$ where $\lceil x \rceil$ denotes the smallest integer that exceeds $x$.  

For fields of order a few Tesla, $\ell_B\approx 10$'s of nm, whereas $k_0$ is expected to be an atomic scale distance of order $0.1\AA^{-1}$.  Hence, quantum oscillations should be observable in slabs a few hundred nm, not too stringent a requirement.  Another practical issue is that of impurities.  Observation of coherent quantum oscillations requires electrons to complete a magnetic orbit before scattering off an impurity: $\omega_c\tau\gg 1$ where $\omega_c \equiv \frac{evB}{2k_0}$ and $\tau$ is the elastic scattering time.  Together with the condition that $B<B_\text{sat}$ this requires that the sample thickness not greatly exceed the mean-free path: $L\ll \ell \equiv \frac{v}{\tau}$.

For $B>B_\text{sat}$ the majority of the magnetic orbit takes place in the bulk, and the energy levels saturate to the field independent values: $\e_n(B\gg B_\text{sat})\approx \frac{\pi v n}{L}$.  Since, the bulk level structure of a thermodynamically thick slab cannot depend on particular choice of boundary conditions, to understand this result, it is useful to compare to a large system with periodic boundary conditions in $y$. There we expect two-sets of $B$-independent energy modes associated with the $\pm$ bulk chiral LL's, each with quantized energies $vk_{y,n} = \frac{2\pi n}{L}$.  
The factor of two difference between the quantization scale periodic and open boundary conditions can be understood as follows: in the periodic boundary-condition case, the chiral LL's from the $\pm$ Weyl nodes are independent leading to a doubly degnerate tower of modes.  In the finite slab, the $\pm$ chiral LL's are no longer separately quantized, since an electron propagating in the `+' Weyl nodes necessarily reflects into the counter-propagating chiral LL of the `-' Weyl node via a detour through the surface states.  However, in the $L\rightarrow \infty$ limit, such distinctions become un-important.  As essential ingredients for the existence of bulk chiral LLs in a thick but finite slab, these magnetic orbits must persist to all energies where the bulk chiral LLs are present.  This expectation is indeed born out by numerical simulations (see below and App.~C of \cite{Supplement}).

\vspace{4pt}\noindent{\bf Direction Dependence of Field}
So far we have considered field along the y-direction, normal to the surface.  For generic field direction $\v{B}$.  The bulk chiral LL's propagate parallel to the field, and only the y-component of $\v{B}$ drives motion of $\v{k}$ along the arc.  Ignoring arc-curvature, one can then simply replace $L\rightarrow \frac{L}{|\hat{y}\cdot\hat{B}|}$ and $k_0\rightarrow \frac{k_0}{|\hat{y}\cdot\hat{B}|}$ in Eqs.~\ref{eq:SemiClassicalQuantization},\ref{eq:Binv}.  Notably, the field-scales $\frac{1}{B_n}$ of Eq.~\ref{eq:Binv} extrapolate to the residual $n=0$ value: $\frac{e\pi v_\perp \gamma}{k_0\mu}|\hat{y}\cdot\hat{B}|+ek_0^{-1}L$.  The constant piece, $eL/k_0$, encodes the deviation of the orbit from a purely 2D orbit and can be extracted by fitting the angle or energy dependence.

\vspace{4pt}\noindent{\bf Deviations from Adiabaticity - } In the above treatment, we assumed that electrons slide all the way to the end of the surface arc before transitioning into the bulk.  A more careful analysis (see App.~B of \cite{Supplement}) shows that the electron jumps off the arc before reaching the end when its momentum is within $\approx \ell_B^{-1}$ of the bulk Weyl node.  This amounts to replacing $k_0\rightarrow k_0-\alpha \ell_B^{-1}$ in Eq.~\ref{eq:SemiClassicalQuantization}, where $\alpha$ is a numerical constant of order unity.  This effect is negligible at low fields $k_0\ell_B \gg 1$, and gives a fractional correction in the quantum oscillation period $\delta\Omega_{B^{-1}}\approx \frac{1}{k_0\ell_B}$ for high fields. The high-field stretching of $\delta\Omega_{B^{-1}}$ is maximal near $B\approx B_\text{sat}$ where $\delta\Omega_{B^{-1}}\approx \frac{1}{\sqrt{k_0 L}}$.

\begin{figure}[tb]
\begin{center}
\includegraphics[width = 3.5in]{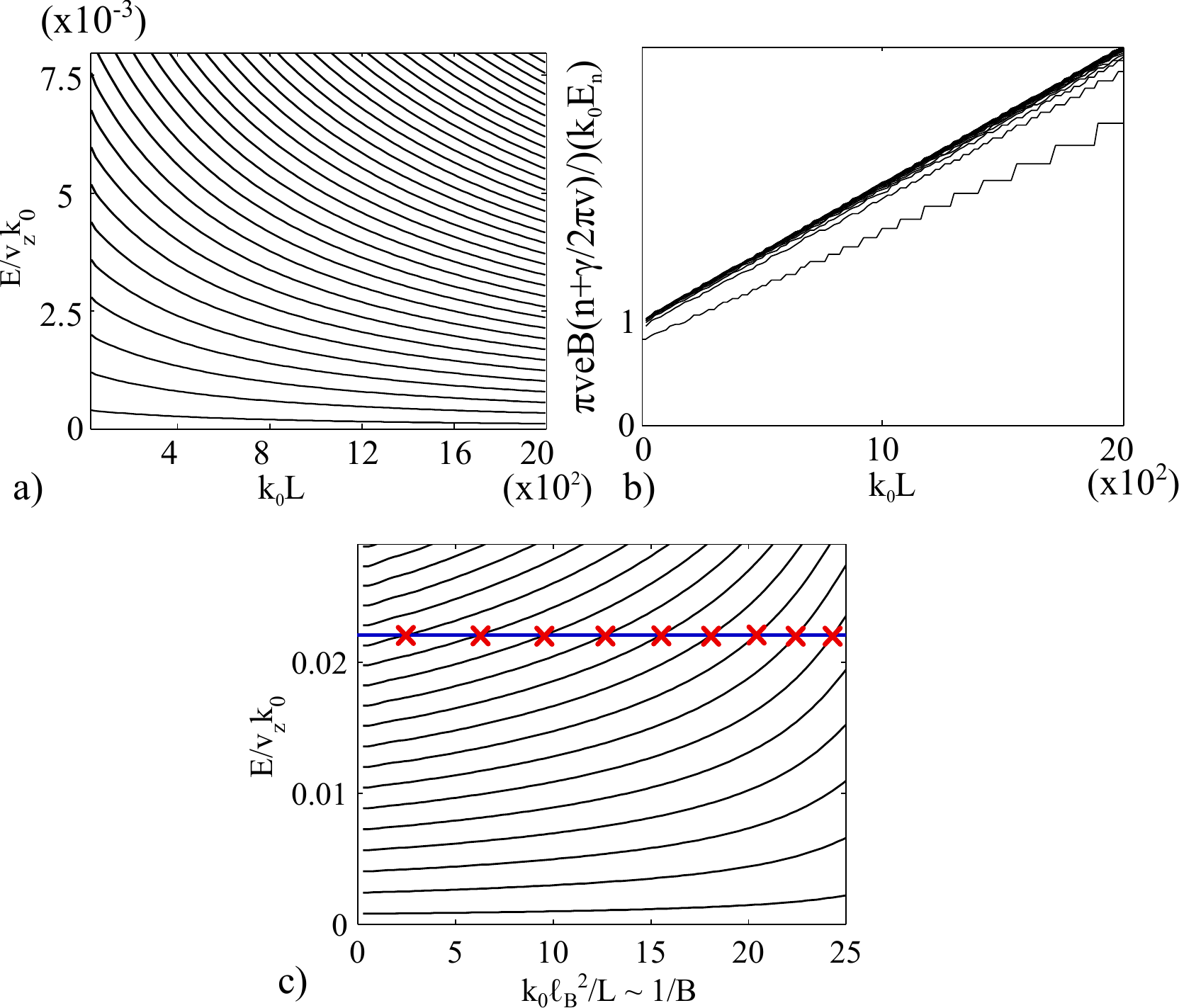}
\end{center}
\vspace{-.2in}
\caption{Energy levels with partial surface-bound state character, obtained from numerical solution of Eq.~\ref{eq:HModel}.  (a) shows bare energy levels for parameters: $v_\perp = 0.2v_z$, and $\ell_B = 28k_0^{-1}$ (corresponding to $B\approx 1$T for typical expected value of $k_0 \approx 0.1\AA^{-1}$).  (b) shows that $n>0$ levels collapse to expected semiclassical form.  Only the lowest energy level deviates from the expected semiclassical form, (the jaggedness of this line is a numerical artifact).  (c) shows $\frac{1}{B}$-dependence of energy-levels with $k_0 L = 15$.  The levels cross (red $\times$'s) a fixed reference energy (horizontal blue line), with nearly equal spacing for $\frac{k_0\ell_B^2}{k_0}\gg 1$ (corresponding to periodic-in-$1/B$ quantum oscillations), then stop crossing for $\frac{k_0\ell_B^2}{k_0} \ll 1$.
}
\vspace{-.2in}
\label{fig:Numerics}
\end{figure} 

\vspace{4pt}\noindent{\bf Numerical Validation of Semiclassical Approximations - }
We validate the semiclassical treatment by directly obtaining the spectrum of a Weyl semi-metal slab in a magnetic field, using the simple model Hamiltonian:
\begin{align} 
H = -iv_\perp\partial_y\sigma^y-\frac{1}{8}v_zk_0\[1-\(2\hat{\pi}_z/k_0\)^2\]\sigma^z+v_\perp \hat{\pi}_x\sigma^x \label{eq:HModel}
\end{align}
(see App.~C of \cite{Supplement} for details). Here, $\boldsymbol{\sigma}$ are $2\times 2$ Pauli matrices labeling two bulk bands, $\hat{\bs{\pi}} = -i\bs{\nabla}-e\v{A}$, $\v{A} = Bz\hat{y}$ is the vector potential in the Landau gauge corresponding to a uniform magnetic field along $\hat{y}$, and we have introduced different velocities $v_z$ along $z$ and $v_\perp$ in the $xy$-plane for convenience.  



Representative results are shown in Fig.~\ref{fig:Numerics}a as a function of slab-thickness $L$.  The curves collapse (Fig.~\ref{fig:Numerics}b) to the predicted semi-classical form of Eq.~\ref{eq:SemiClassicalQuantization}.  Fig.~\ref{fig:Numerics}c shows the energy levels and their intercept with a line of fixed energy as a function of inverse-field.  The crossings are regularly spaced for small fields, and saturate to $B$-independent values at large fields.  For moderate fields, we observe the predicted stretching of the $\frac{1}{B}$-period, due to departures from adiabaticity.  The good agreement between semiclassics and numerics validates the semiclassical approach, which we now extend to treat quantum oscillations from surface states in Dirac materials.



\vspace{4pt}\noindent{\bf Dirac Surface States - }
Consider the surface of a Dirac SM slab with normal along the y-direction. Such a surface breaks the crystal symmetry protecting the bulk Dirac nodes, but let us first ignore this effect.  Then, there will be two sets of Fermi-arcs arising from the two copies of the bulk Weyl nodes, denoted R and R', that transform differently under crystal symmetry and generically curve in opposite directions.  The surface arcs join at the bulk Weyl nodes without mixing leading to a sharp corner or ``kink" in the shape of the surface-states' Fermi-surface (Fig.~\ref{fig:DiracCase}a).  Such a kink would not be allowed in a purely 2D metal, and is a special feature of the crystal-symmetry-protected Weyl structure of the Dirac SM.  

Next, imagine adding back the crystal symmetry-breaking effects of the surface.  The surface states are clearly stable to a weak crystal symmetry-breaking potential away from the Dirac nodes where the two surface arcs are separated in momentum. Interestingly, the perturbative protection extends all the way to the bulk Dirac node, since the surface wave-function becomes extended into the bulk near the bulk Dirac-points and is no longer affected by the surface (see App. D of \cite{Supplement}).  Therefore, when the symmetry-breaking potential of the surface is not too strong, the surface arcs survive and their Fermi-surface retains the unconventional ``kink"-discontinuity.  On the other hand, a sufficiently strong surface potential can reconstruct (Fig.~\ref{fig:DiracCase}b,c) or even entirely remove (Fig.~\ref{fig:DiracCase}d) the surface states. 

In the following section, we consider the case where the surface states are connected to the bulk Dirac-points (Fig.~\ref{fig:DiracCase}a or b), and ask whether these unconventional kinked Fermi-surfaces lead to quantum oscillations in a magnetic field.

\vspace{4pt}\noindent{\bf Quantum Oscillations from Dirac Surface States - }
For Cd$_3$As$_2$ and Na$_3$Bi, an applied magnetic field along any axis besides $z-$ breaks the discrete rotation symmetry protecting the Dirac nodes.  Consequently, the bulk chiral LLs for the two sets of overlapping Weyl nodes are mixed by the field, and develop a gap $\Delta_B\approx CeB\[1-(\hat{B}\cdot\hat{z})^2\]$, where $C$ is a material parameter (see App. E of \cite{Supplement}).  Henceforth, we will refer to this effect as RR'-mixing. States with energies $\e>\Delta_B$ are unaffected by this gap, and the magnetic orbits proceed as two independent copies of the Weyl orbits.  

For $\e<\Delta_B$, the RR'-mixing can back-scatter electrons between segments $2\leftrightarrow 3'$ and $3\leftrightarrow 2'$ of the orbits in Fig.~\ref{fig:DiracCase}a, characterized by the length scale $\ell_M \approx \frac{v}{2\Delta_B}$.  If the distance travelled through the bulk is much less than $\ell_M$, i.e. $\frac{L}{|\hat{B}\cdot\hat{y}|}\ll\ell_M$, then the RR'-mixing does not typically occur before the electron traverses the bulk.  In this case, the mixing is ineffective and the magnetic orbits for each Weyl copy occur essentially independently. 

On the other hand, if $\frac{L}{|\hat{B}\cdot\hat{y}|}\ll\ell_M$, the electrons penetrate only distance $\ell_M$ into the bulk before returning to the same surface (but switching copies).  Then, the orbits occur separately on each surface.  The time to traverse such a semi-classical orbit is $t\approx \frac{2k_0}{ev(\v{B}\cdot\hat{y})}+\frac{2\ell_M(\v{B})}{v}$
where the second term accounts for the time virtually spent in the bulk chiral LLs. Applying the semi-classical quantization condition leads to the corresponding energy levels:
\begin{align} \e_n = (n+\gamma) \frac{eB}{m_c^\text{eff}(\hat{B})}\label{eq:SpectrumThickLimit}
\end{align}
where we have defined an effective cyclotron mass:
\begin{align} m_c^\text{eff}(\hat{B}) = \frac{2k_0}{\pi v|\hat{B}\cdot\hat{y}|}+\frac{1}{2\pi C[1-(\hat{B}\cdot\hat{z})^2]} \label{eq:EffCyclotronMass}\end{align}
The density of states at fixed energy, $\mu$, has periodic-in-$1/B$ oscillations with period $\Omega_{B^{-1}} = e/\mu m_c^\text{eff}$

\vspace{4pt}\noindent{\bf Evolution of Quantum Oscillations as a Function of Field Strength - }
In a typical quantum oscillations experiment, one maps the density of states a fixed reference energy $\mu$, $\nu(\mu)$, as a function of inverse field strength, $\frac{1}{B}$.  For small $B$ ($\ll B_\text{sat,mix}$), the R and R' orbits are isolated and quantum oscillations occur with period corresponding to Eq.~\ref{eq:SemiClassicalQuantization}.  For large field, $B\gg B_\text{mix}$, and $\e\ll \Delta_B$, the copies are mixed and the orbits occur separately on top and bottom surfaces.  The detailed crossover behavior depends on two characteristic field scales: 1) the field scale at which the energies of the isolated orbits $1234$ and $1'2'3'4'$ saturate and become roughly independent of field: $\frac{1}{eB_\text{sat}} = L/k_0$ and 2) the characteristic scale at which the slab thickness $L$ exceeds the inter-copy mixing scale $\ell_M$: $\frac{1}{eB_\text{mix}} = \frac{CL[1-(\hat{B}\cdot\hat{z})^2]}{v|\hat{B}\cdot\hat{y}|}$.  Denote the ratio of these to scales by: $\beta \equiv \frac{B_\text{sat}}{B_\text{mix}}$. 

If $\beta <1$, there is a re-entrant behavior where quantum oscillations occur for $B<B_\text{sat}$ and $B>B_\text{mix}$, but cease in the intervening regime $B_\text{sat}<B<B_\text{mix}$ where energy levels have only weak field dependence.    If $\beta>1$, $B_\text{mix}$ pre-empts $B_\text{sat}$, and there are quantum oscillations for all fields. At large fields, $B\gg B_\text{mix}$, there are two sets of identical energy levels on the top and bottom surface (corresponding to semiclassical orbits $11'$ and $33'$ of Fig.~\ref{fig:DiracCase}a) whose energies are given by Eq.~\ref{eq:SpectrumThickLimit}.  As $B$ is decreased through $B_\text{mix}$, the $11'$ and $33'$ orbits hybridize through the bulk and split with scale $\Gamma\sim e^{-2L/\ell_M}$. For $B\ll B_\text{mix}$, the levels recombine into those associated with the two isolated orbits $1234$ and $1'2'3'4'$ (see Fig~\ref{fig:DiracCase}a).  The crossover can be continuously probed in a single sample by varying the field-direction.

\vspace{4pt}\noindent{\bf Discussion - }  To summarize, by a combination of semiclassical analysis and direct numerical solution, we have demonstrated that up to a critical field-strength that is inversely proportional to sample thickness, the surface Fermi-arcs of Weyl semi-metals contribute periodic-in-$1/B$ quantum oscillations in the density of states.  Similar quantum oscillations can occur from arc-like surface states in Dirac semi-metals.  In both cases, density of states oscillations in a field provide experimentally testable fingerprints of the unconventional Fermiology of the Weyl and Dirac surface-states.  


\noindent\emph{Acknowledements}
We thank S.A. Parameswaran, P. Hosur, J.P. Dahlhaus, Y.-M. Lu, and A. Yazdani for helpful conversations. A.V. was funded by NSF DMR 0645691.

\onecolumngrid
\newpage
\appendix
\large

\section{\large Appendix A. Spectrum of a Semi-Infinite Weyl SM Slab \label{App:Spectrum}}
A slab of Weyl SM, infinite in the x- and z-directions, and semi-infinite in the y-direction, filling the $y>0$ half-plane, can be modeled by Eq.~\ref{eq:HModel}.

For the purpose of these appendices, it will be sufficient to linearize the dispersion in the vicinity of a single Weyl node.  For definiteness and simplicity, we show the results only for a single chirality Weyl node, ignore any velocity anisotropy and choose units where the Weyl velocity $v$ is unity (factors of $v$ can easily be restored later by dimensional considerations).  With these simplifications, the linearized Hamiltonian is:
\begin{align} H = -i\partial_y\sigma^y+k_x\sigma^x+k_z\sigma^z+M(y)\sigma^z \end{align}
where $M(y) = M$ for $y<0$, $M(y)=0$ for $y>0$, and we take implicitly work in the limit $M\rightarrow \infty$ to model the interface with vacuum or a large-gap trivial insulator.

Appropriate boundary conditions at $y=0$ are determined by solving for eigenstates of $H$ in the $y>0$ regime, and then taking $M\rightarrow \infty$.  The results in the boundary condition:
\begin{align} \Psi(y=0)\sim \begin{pmatrix} 1\\ 1\end{pmatrix} \end{align}

There are surface states for $k_z<0$ with wave-function:
\begin{align} \Psi_0 = \sqrt{|k_z|}e^{ik_xx}e^{-|k_z|y}\frac{1}{\sqrt{2}}\begin{pmatrix}1\\ 1\end{pmatrix}
\end{align}
with energy $k_x$.  In What follows, we will consider only $k_x=0$ for simplicity.

To obtain the spectrum and wave-functions of extended bulk states, it is useful to re-write the Hamiltonian (with $k_x=0$) as:
\begin{align} H = E(\cos\theta\sigma^y+\sin\theta\sigma^z)
\\ 
\theta\equiv \tan^{-1}p/k_z\end{align}
The eigenstates with energy $\pm E$ are:
\begin{align} \Psi_\pm(p) \simeq e^{ipy} \begin{pmatrix} \cos(\theta_p/2)\pm \sin(\theta_p/2) \\ i\[\pm \cos(\theta_p/2)-\sin(\theta_p/2)\] \end{pmatrix}\end{align}

We need to superpose $\Psi_\pm(p)$ with $\Psi_\pm(-p)$ to match the boundary conditions at $y=0$.  For example, for $\Psi_+$:
\begin{align} \alpha_p\Psi_+(p,y=0)+\alpha_{-p}\Psi_+(p,y=0) &= \begin{pmatrix} \(\alpha_{p}+\alpha_{-p}\)c+\(\alpha_{p}-\alpha_{-p}\)s
\\
i\(\alpha_{p}+\alpha_{-p}\)c-i\(\alpha_{p}-\alpha_{-p}\)s
\end{pmatrix}\sim \begin{pmatrix} 1\\ 1\end{pmatrix} 
\end{align}
which requires: $\alpha_{\pm p} = \frac{1}{\sqrt{2L_y}} e^{\pm i\theta_p/2}$.  Here $L_y$ is the length in the y-direction (to be taken to $\infty$).
\begin{align}\Psi_{+E}(y) = \frac{1}{2\sqrt{L_y}}\[e^{ipy+i\theta_p/2}e^{-i\theta_p/2\sigma^x}\begin{pmatrix}1\\ i\end{pmatrix}+e^{-ipy-i\theta_p/2}e^{+i\theta_p/2\sigma^x}\begin{pmatrix}1\\ -i\end{pmatrix}\]\end{align}
Similar expresssions can be found for the $-E$ solutions.

\section{\large Appendix B. Diabatic Corrections to Semi-Classics}

In this section, we develop a perturbation theory method for systematically computing corrections to adiabaticity.  We would like to estimate how close (in momentum space) an electron sliding along the surface Fermi-arc gets to the bulk Weyl node before transitioning to the bulk.  Denote this quantity by $\delta k(B)$.  Before entering a detailed calculation, it is useful to note that, for $k_0\ell_B\gg 1$, the magnetic field can induce transitions from surface Fermi-arcs to bulk states only very near the bulk Weyl node.  Close to the bulk Weyl node, the zero-field spectrum is scale invariant, and the only lengthscale in the problem is $\ell_B$.  Hence, by dimensional considerations, the only possible answer is that $\delta k(B)$ is equal to $\ell_B^{-1}$ times a universal constant.  Indeed, a more detailed calculation below will confirm this general argument.

\subsection{Time-Evolution in the ``Adiabatic Picture"}
For a given time-dependent Hamiltonian $H(t)$, it is natural to define the ``adiabatic picture" (A-picture) eigenstates by absorbing the adiabatic part of the time-evolution into the Heisenberg picture (time-independent) eigenstates of $H(t=0)$:
\begin{align}|\Psi(t')\>_A \equiv \(\sum_n|\Psi_{i,n}(t)\>\<\Psi_{i,n}(t)|\)|\Psi(t)\>_A\equiv U_i(t',t)|\Psi(t)\>_A
\end{align}
where $\mathcal{T}$ indicates time-ordering, and $\{|\Psi_{i,n}(t)\>\}$ are the instantaneous eigen-states of $H(t)$ (i.e. the eigenstates of the time-independent Hamiltonian $H'\equiv H(t)$).  This time-evolution would be exact in the adiabatic limit, and the A-picture simplifies the computation of perturbative corrections for nearly adiabatic evolution.

The time-evolution operator in the A-picture becomes:
\begin{align} U_A(t',t) &= U_i(t',0)U(t',t)U_i(t,0)^\dagger = \lim_{N\rightarrow \infty} \prod_{j=1}^N U_i(t_j+\delta t,0)U(t_j+\delta t,t_j)U_i(t_j,0)^\dagger 
\nonumber\\
&= \lim_{N\rightarrow \infty} \prod_{j=1}^N U_i(t_j+\delta t,0)\[1-iH(t_j)\delta t\]U_i(t_j,0)^\dagger 
\nonumber\\
&= \lim_{N\rightarrow \infty} \prod_{j=1}^N U_i(t_j+\delta t,0)U_i(t_j,0)^\dagger +U_i(t_j,0)\(-iH(t_j)\delta t\)U_i(t_j,0)^\dagger
\nonumber\\
&= \lim_{N\rightarrow \infty} \prod_{j=1}^N 1-i\delta t\[-iU\partial_t U^\dagger+U_i(t_j,0)H(t_j)U_i(t_j,0)^\dagger\]
\equiv \mathcal{T}e^{-i\int^t H_A}\end{align}
where:
\begin{align} H_A(t) \equiv -iU\partial_t U^\dagger+U_i(t,0)H(t)U_i(t,0)^\dagger \end{align}

For adiabatic evolution, the first term is vanishingly small, and the second-term is simultaneously diagonalized by the same states for all times.  The leading-order diabatic correction may be obtained by treating the first term as a (time-dependent) perturbation in the usual way, by moving to the interaction picture with respect to $U_i(t,0)H(t)U_i(t,0)^\dagger$.

\begin{align}H_\text{dia}=-iU\partial_t U^\dagger = \lim_{t'\rightarrow t} U(t',t) = \lim_{t'\rightarrow t}  \(\sum_n i\partial_{t'}|\Psi_{i,n}(t')\>\<\Psi_{i,n}(t)|\)\end{align}
In the basis of instantaneous eigen-states of $H(t)$:
\begin{align} \[H_\text{dia}(t)\]_{nm} = i\<\partial_t\Psi_{i,n}(t)|\Psi_{i,m}(t)\> = i\sum_{m\neq n}\frac{\<\Psi_{i,n}(t)|\dot{H}(t)|\Psi_{i,m}(t)\>}{E_n-E_m} \end{align}
where we have assumed a gapped spectrum to expand instantaneous eigen-states at $t+\delta t$ in terms of those at $t$ perturbatively in $\partial_tH(t)\equiv \dot H$.

\begin{align} \[H_A(t)\]_{n,m} = \sum_{m\neq n}\frac{\<\Psi_{i,n}(t)|\dot{H}(t)|\Psi_{i,m}(t)\>}{E_n-E_m} +E_n(t)\delta_{n,m}
\end{align}

\begin{align}  \[H_\text{dia}(t)\]_{\text{int};n,m}\equiv e^{i\int^t(E_n-E_m)} \[H_A(t)\]_{n,m} \end{align}

The amplitude for inducing a diabatic transition from an initial instaneous eigenstate $0$ to a different state $m$, to leading order in $H_\text{dia}$ is:
\begin{align} c_m^{(1)}(t) = -i\int_0^t dt_1\[H_\text{dia}(t_1)\]_{\text{int};m,0} \label{eq:TransitionAmplitude}\end{align}
and the probability of any diabatic transition is then:
\begin{align} P(t)= \sum_{m\neq 0} |c_m^{(1)}(t)|^2 \end{align}

\subsection{Application to Weyl Semi-Metal in a field}
The effect of a magnetic field forcing an electron to slide along a Fermi-arc can be roughly modeled by taking to $k_z$ be time-dependent: 
\begin{align} k_z(t) = \dot{k}t \end{align}
with $\dot{k} = evB=v\ell_B^{-2}$.  For simplicity, let us choose a guage in which $k_x$ is a good quantum number, and restrict our attention to $k_x=0$.  Henceforth, we will ignore anisotropy in velocity and choose units where $v=1$.

\begin{figure}
\begin{center}
\includegraphics[width = 1.5in]{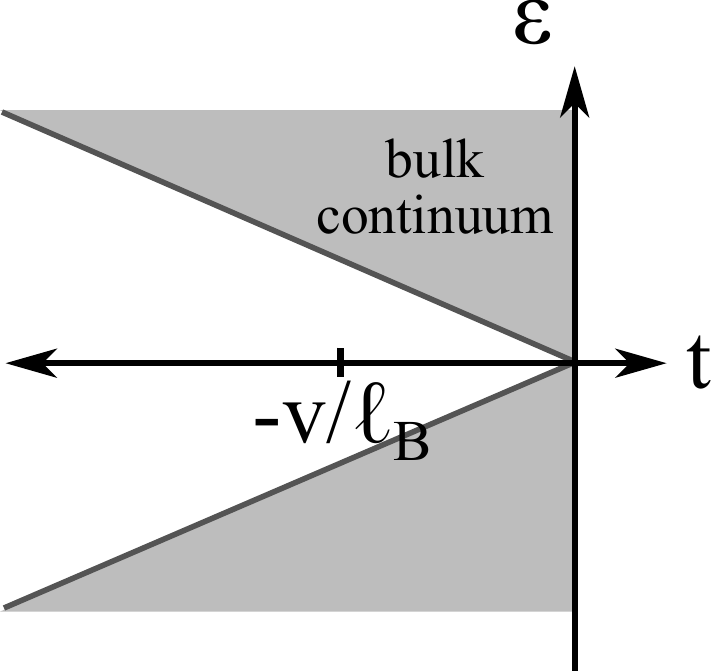}
\end{center}
\vspace{-.2in}
\caption{Schematic of instantaneous eigenstates for momentum sliding along the surface-Fermi arc according to $k_z(t)=evB$.  The particle typical transitions to the bulk at time of order $\frac{v}{\ell_B}$ before reaching the end of the arc. 
}
\vspace{-.2in}
\label{fig:AppFig1}
\end{figure} 

Using the results from Appendix~A, the relevant quanties for computing the transition probability $P(t)$ are:
\begin{align} E_p(t) = \sqrt{p^2+k_z(t)^2} \end{align}
\begin{align} \varphi(t)\equiv \int^t E_p &= \frac{1}{\dot{k}}\int_{k(t)}^0 dk \sqrt{p^2+k^2} = \frac{1}{\dot{k}}\[k E_p+p^2\log\(k+E_p]\)\]|_{k(t)}^0
\nonumber\\
&=
\frac{1}{\dot{k}}\[p^2\log\(\frac{p}{k+E_p}\)-kE_p\](t)
\end{align}

\begin{align} \dot H(t) = \dot k_z(t)\sigma^z = \dot{k}\sigma^z\end{align}

\begin{align} \<\Psi_{i,p}(t)|\dot{H}(t)|\Psi_{i,0}(t)\> 
&=\frac{\dot{k}\sqrt{k}}{\sqrt{2}L_y}\int dy \cos\[py+\theta_p/2-\pi/4\]e^{-|k|y} 
= \frac{2\dot{k}\sqrt{k}}{\sqrt{2}L_y}\cos\[\frac{3\theta_p}{2}-\pi/4\]
\end{align}

Combinging these expressions into Eq.~\ref{eq:TransitionAmplitude} gives:
\begin{align} c_p^{(1)}(t) \approx \frac{2}{\sqrt{2}L_y}\int_{-\infty}^{k(t)}dk e^{i\varphi(k)}\frac{\sqrt{|k|}}{\(k^2+p^2\)}\cos\(3\theta_p/2-\pi/4\)
\end{align}

The dominant contributions to the time integral in $c_m^{(1)}$ come from $p\ll k$. For $k(t)>\sqrt{\dot{k}}$, when the electron is far from the bulk Weyl node and the bulk gap is large, the phase factor $e^{i\varphi(t)}$ is rapidly oscillatory, and the probability of diabatic transition is exponentially small ($\sim e^{-k(t)^2/\dot{k}}$).  As $k(t)$ becomes small the bulk gap closes and the transition probability diverges logarithmically:
\begin{align} P(t) \approx \log\frac{\dot{k}}{k(t)} \end{align}
$P(t)$ necessarily becomes of $\mathcal{O}(1)$ when $k(t)\lesssim \dot{k}/v = \ell_B^{-1}$.

This indicates that the magnetic field causes an electron, initially on the surface Fermi-arc, to transition into the bulk states when the electron's momenta comes within $\approx\ell_B^{-1}$ of the bulk Weyl node.

\section{\large Appendix C. Details of Numerics}
The Weyl SM slab in a field is modeled by Eq.~\ref{eq:HModel} with $\v{A} = Bz\hat{y}$ is the vector potential in the Landau gauge corresponding to a uniform magnetic field along $\hat{y}$.  In this gauge, x-momentum is a good quantum number, which just sets the location of the guiding center for the Landau orbits, and can be taken as zero without loss of generality.  Then, we introduce Landau-level raising and lowering operators via: 
\begin{align} \pi_x = \frac{1}{\sqrt{2}\ell_B}\(a^\dagger+a\)
~~~~~~
\pi_z = \frac{-i}{\sqrt{2}\ell_B}\(a^\dagger-a\) 
\end{align}
which satisfy canonical commutation relations $\[a,a^\dagger\]=1$.  It is convenient to divide Eq.~\ref{eq:HModel} into anti-commuting pieces,  $H=H_y+H_\perp$ where:
\begin{align} 
H_y&\equiv -iv_\perp\partial_y\sigma^y
\nonumber\\
H_\perp &\equiv -\frac{1}{8}v_zk_0\[1-\(2\hat{\pi}_z/k_0\)^2\]\sigma^z+v_\perp \hat{\pi}_x\sigma^x \label{eq:HModel}
\end{align}

Since $H_y$ and $H_\perp$ anti-commute, eigenstates of the full Hamiltonian $H=H_y+H_\perp$ can be constructed from the separate knowledge of the eigenstates and energies of $H_y$ and $H_\perp$.  Eigenfunctions of $H_y$ have the form $\sim e^{\pm ik_y}$.  Real values of $k_y$ correspond to extended plane-wave like states, whereas purely imaginary $k_y$ correspond to surface states bound to either the top- ($\text{Im}~k_y>0$) or bottom- ($\text{Im}~k_y<0$) surface respectively. Eigenstates of $H_\perp$ are obtained numerically, by truncating the Hilbert space to keep only a finite number of Landau-Levels\cite{Endnote:LLTruncation}.  States corresponding to the semiclassical orbit of Fig.~\ref{fig:SemiclassicalTrajectory} are a mixture of extended bulk states with real $k_y$ and surface bound-states with purely imaginary $k_y$.

The states  of the reduced Hilbert space of $H_\perp$ for fixed guiding center position are then labeled by orbital index, e.g. $\sigma^y = \pm 1$, and LL oscillator level $n$ (defined by $a^\dagger a |n\> = n|n\>)$.  The eigenstates of $H_\perp$ are difficult to obtain analytically, but can easily be obtained numerically, by truncating the Hilbert space to keep only a finite number of $|n\>$ with $n<N_\text{max}$\cite{Endnote:LLTruncation}. Let us denote the positive energy eigenstates of $H_\perp$ by $\begin{pmatrix}f_m(x,z)\\ g_m(x,z)\end{pmatrix}$ (in the basis where $\sigma^y$ is diagonal), having energy $\e_{\perp,m}>0$.  Since $\sigma^yH_\perp\sigma_y = -H_\perp$, the spectrum of $H_\perp$ is particle-hole symmetric, and hence negative energy eigenstates can be obtained from positive energy ones by flipping the sign of $g$.

Even though $k_z$ is no longer a quantum number in the presence of a magnetic field, the magnetic-field effectively smears $k_z$ by size $\ell_B^{-1}$, and for $k_0\ell_B\gg 1$, the bulk Weyl nodes are well isolated from one another.  In this case, $H_\perp$ will also have zero-mode solutions, corresponding to the bulk chiral $n=0$ LLs.  These are necessarily $\sigma^y$ eigenstates due to the particle-hole symmetry of $H_\perp$.  

Eigenfunctions of $H_y$ are easily identified as $e^{isk_y}|\sigma_y = s' = \pm 1\>$, corresponding to eigenvalue $\e_y = ss'v_\perp k_y$.  Real values of $k_y$ correspond to extended plane-wave like states, whereas purely imaginary $k_y$ correspond to surface states bound to either the top- ($\text{Im}~k_y>0$) or bottom- ($\text{Im}~k_y<0$) surface respectively.

Since $H_y$ and $H_\perp$ anti-commute, eigenstates of the full Hamiltonian $H=H_y+H_\perp$ can be constructed from the separate knowledge of the eigenstates and energies of $H_y$ and $H_\perp$.  Specifically, the eigenstates of $H$ have particle-hole symmetric spectrum with energies: $E_{k_y,m}^{(\pm)} = \sqrt{\e_y^2+\e_{\perp,m}^2}$.  For $\e_\perp\neq 0$, there are four corresponding eigenstates of the form:
\begin{align} \Psi_{E,k,m}^{(s,s')} = e^{isv_\perp k}\begin{pmatrix}\(E+sv_\perp k\)f_m(x,z)\\ s'\e_{\perp,m}g_{m}(x,z)\end{pmatrix}\end{align}
where $s,s'=\pm$.  

Only certain superpositions of such eigenstates will satisfy the  boundary conditions at the bottom ($y=0$) and top ($y=L$) surfaces of the slab.  The proper boundary conditions can be identified by treating the vacuum ($y<0$ and $y>L$) as an ordinary band-insulator with large gap.  Equivalently, we can add a mass term to Eq.~\ref{eq:HModel} of the form $M(y)\sigma^z$, where $M(y) = 0$ inside the slab, and $M(y) = M_0>0$ outside the slab.  Continuity of the wave-functions at the slab-boundaries, upon taking the limit $M_0\rightarrow \infty$, require that $\Psi(y=0,L)$ be eigenstates of $\sigma^x$ with eigenvalues $\pm 1$ respectively.

\begin{figure}
\begin{center}
\includegraphics[width = 4in]{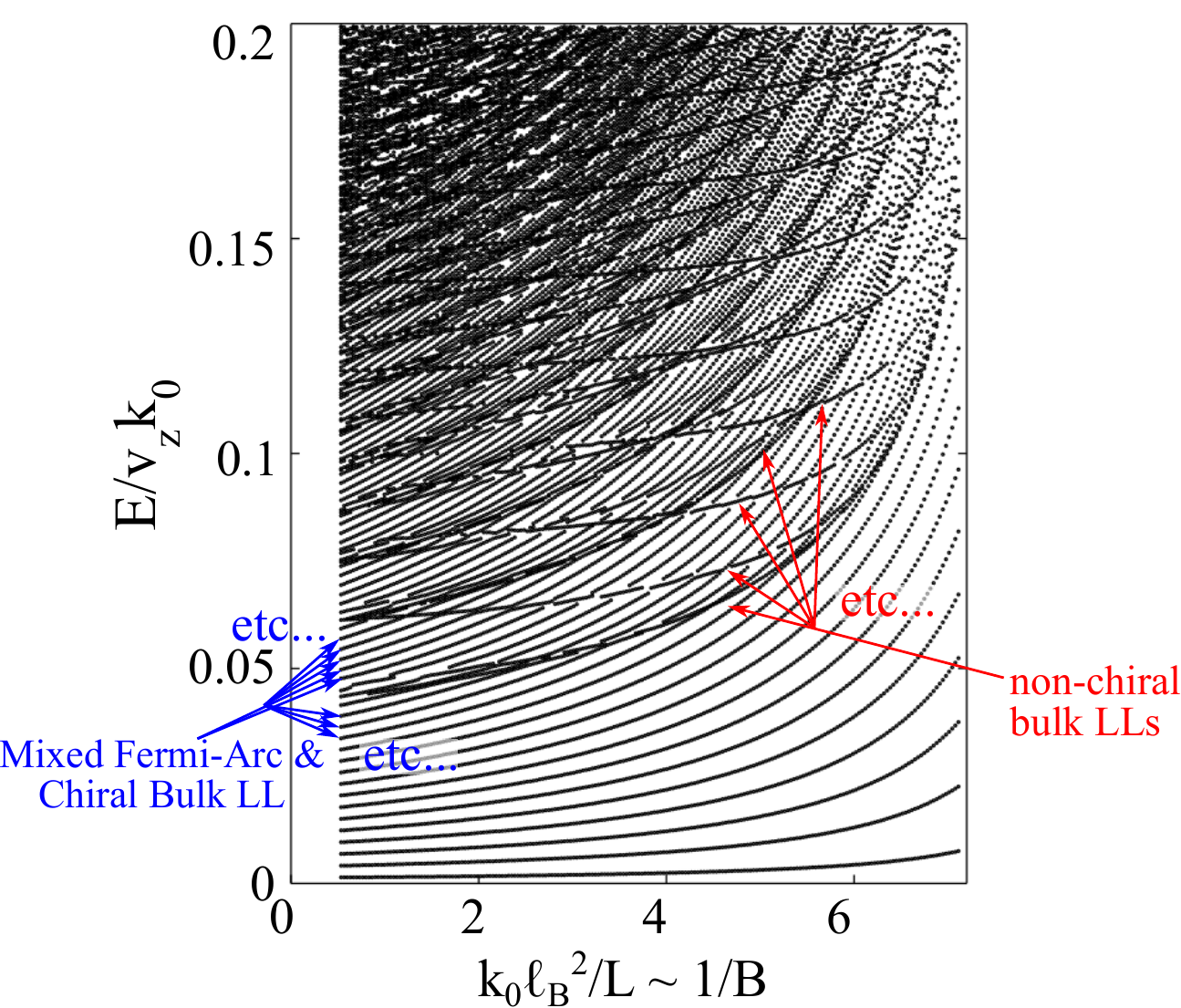}
\end{center}
\vspace{-.2in}
\caption{Numerically obtained spectrum for a Weyl semi-metal slab with a wider range of energy above the bulk Weyl node for the same parameters as Fig.~\ref{fig:Numerics}c in the main text, except for slab thickness $L = 30/k_0$.  The levels with mixed surface-arc and bulk-chiral LL character exist up to higher energies (examples are indicated by blue arrows), but are also accompanied by states from the non-chiral bulk LLs (examples are indicated by red arrows).
}
\vspace{-.2in}
\label{fig:AppFig4}
\end{figure}

As usual, these boundary conditions can only be satisfied for a discrete set of energies, $E$, by superposing approriate linear combinations of $\Psi_{E,k,m}^{(s,s')}$. Extended states, with real $k$ and $\e_\perp\neq 0$ can be obtained by the superposition $\Psi_{E,k,m} = \sum_{s,s'}(-1)^{ss'}\Psi_{E,k,m}^{(s,s')}$, subject to the condition $k = \frac{2\pi n}{L}$ with $n\in \mathbb{Z}$.  Such extended states are associated with the gapped bulk LLs with LL index $|n|>0$.

The chiral bulk LL's with $\e_\perp = 0$ cannot, by themselves, be superposed to match the required boundary conditions, since appearance of zero modes in $H_\perp$ requires good isolation between the $\pm$ Weyl nodes.  Under these conditions the upward moving $n=0$ LL from the `+' Weyl node cannot reflect directly into the opposite downward moving mode from the `-' Weyl node.  However, we can construct low-energy eigenstates that are mixtures of the $\e_\perp=0$ extended states and surface bound-states with purely imaginary $k$ and $\e_\perp >0$:
\begin{align} \Psi_E=\sum_{s,m}\gamma_{s,m}e^{-sy\kappa_m} \begin{pmatrix} (E+is\kappa_m)f_m \\ |\e_{\perp,m}| g_m \end{pmatrix}
\end{align} 
where $s=\pm 1$, $m>0$, and $\kappa_m = \sqrt{\e_{\perp,m}^2-E^2}$.  Such states are precisely those described by the semiclassical orbit of Fig.~\ref{fig:SemiclassicalTrajectory}.   The boundary condition matching condition for these hybrid surface/bulk orbits can be written as a homogenous linear equation for $\gamma_{s,m}$ of the form: $\hat{M}(E)\gamma = 0$ where $\hat{M}(E)$ is a matrix that depends on $E$.  Allowed energies, $E$, can be found by numerically varying $E$ to search for a vanishing singular value of $\hat{M}(E)$.

Some representative results are shown for energies below that of the lowest non-chiral bulk LL in Fig.~\ref{fig:Numerics} of the main text.  Fig.~\ref{fig:AppFig4} above shows a broader range of energies. By inspection, one can clearly see the levels corresponding to mixed surface-arc and bulk chiral LL orbits persist to higher energies, coexisting with the bulk non-chiral LL's.  The states involving surface-arcs will have larger weight compared to bulk levels on the surface, and are expected to be more visible in surface-sensitive probes like tunneling (as opposed to probes that are not surface selective like Shubnikov-de Haas oscillations of electrical conductivity).

\section{\large Appendix D. Dirac Semi-Metal Surface States}
Consider a surface of a Dirac SM with normal in the $y$-direction (the analysis can be generalized to more generic surfaces, with similar results, unless the surface-normal lies exactly along the $z$-axis connecting the two different valleys in the bulk).  After projecting into the low-energy, long-wavelength manifold of the Dirac nodes, the Hamiltonian can be written as:
\begin{align} 
H &= H_0+ V_\text{surface}+V_\text{Bulk-Curvature}
\nonumber\\
H_\text{0}&=v\v{p}\cdot\bs{\tau}\sigma^z\eta^z+M\theta(y)\sigma^z\tau^z
\nonumber\\
V_\text{surface} &=  \delta(y) V_{abc}\tau^a\sigma^b\eta^c
\nonumber\\
V_\text{Bulk-Curvature} &= V_{BC} p^z\eta^z
\end{align}
where $\tau$, $\sigma$, and $\eta$ are Pauli matrices labeling pseudo-spin, crystal-symmetry representation, and valley. Here, for simplicity, we model the vacuum by a large-gap trivial insulator, corresponding to the term $M\theta(y)\sigma^z\tau^z$ (with $M$ implicitly taken to be very large).  $V_{abc}$ models the effects of the surface potential.  Without loss of generality, is non-zero only for values of $abc$ which 1) anticommute with the y-dispersion $\sigma^y\sigma^z$ (since we have chosen the surface with $y$-normal valley is a good quanum number and $V_\text{surface}$ may not contain $\tau^{x,y}$) and 2) preserve the residual symmetries of the surface.

Let us divide $V_\text{surface}$ into terms that are purely diagonal (off-diagonal) in the $\tau^z$ eigen-basis, denoted $V_1 \equiv \delta(y)\sum_{b=0,3}M_\text{abc}\tau^a\sigma^b\eta^c$ and $V_2 \equiv \delta(y)\sum_{b=1,2}M_\text{abc}\tau^a\sigma^b\eta^c$ respectively.  

We first consider the eigenstates of $H_0$. $H_0$ does not mix the superposed Weyl SM copies, labelled R and R', and hence has two sets of oppositely oriented surface-arcs with wave-functions $|\Psi_{R,R'}(k_z,k_x)\>$.  The Fermi-arcs for $H_0$ alone lie directly on-top of each-other.  For fixed $k_z$ along the arc, we can model the surface-states of $H_0$ by the Hamiltonian:
\begin{align} H_\text{Arc,0} = v_\text{arc}p_x\gamma^z \end{align}
where $\{\gamma\}$ are $2\times 2$ Pauli matrices, with $\gamma^z$ diagonal in the $|\Psi_{R,R'}(k_z,k_x)\>$ basis, with eigenvalues $\gamma^z=\pm 1$ corresponding to the R and R' arcs respectively.

$V_1$ and $V_\text{Bulk-Curvature}$ do not mix R and R', and hence when projected into the surface arc basis, $|\Psi_{R,R'}(k_z,k_x)\>$, can renormalize the arc velocity $v_\text{arc}$ and introduce a diagonal term $V_C(k_z)$.  $V_C(k_z)$ vanishes linearly in $k_z$ near the two bulk Dirac nodes, $V_C(k_z\approx 0)\approx Bk_z$, since in the vicinity of the Dirac nodes: 1)  $V_\text{Bulk-Curvature}$ vanishes linearly in $k_z$, and 2) the matrix element for any surface potential vanishes linearly in $k_z$ since, $\<\Psi_{R,R'}|\delta(y)\Psi_{R,R'}\>\approx k_z$.

Lastly, projecting the $V_2$ into the surface arc basis, $\<\Psi_R(k_z)|V_2|\Psi_R'(k_z)\> = \Delta(k_z)$.  Again, the matrix elements vanish linearly as $\Delta(k_z)\approx \Delta_0k_z$ near the bulk Dirac nodes.

Combining these elements gives the following effective Hamiltonian for the surface arcs:
\begin{align} H_\text{arc,eff} = v_\text{arc}k_x\gamma^z+V_C(k_z)+\Delta(k_z)\gamma^x \label{eq:Harceff}
\end{align}
where the first term represents the flat surface Fermi-arcs of $H_0$.  The second term causes the R and R' arcs to have opposite curvature and produces a kink near the bulk Dirac nodes, where the arcs join at a sharp corner.  The last term represents inter-arc mixing due to the reduction of crystal symmetries at the surface which tends to gap out the surface arcs.

The unconventional feature of the bare surface arcs is the corner/kink feature where they meet the bulk.  To see if this feature persists in the presence of the surface-term $\Delta$, let us linearize Eq.~\ref{eq:Harceff} near $k_z\approx 0$.  The energies are $E^{\pm}(k_z,k_x) = Bk_z\pm\sqrt{(v_\text{arc}k_x)^2+(\Delta_0k_z)^2}$.  

When $|B|>|\Delta_0|$, the Fermi-surface is parameterized by separate arcs: $k_x = \pm \sqrt{\frac{B^2-\Delta_0^2}{v_\text{arc}^2}}~k_z$, which meet at the bulk nodes with a sharp corner with angle $\alpha = \tan^{-1}\sqrt{\frac{B^2-\Delta_0^2}{v_\text{arc}^2}}$. This situation is shown in Fig.~\ref{fig:DiracCase}a,b of the main text.  If $|V_C(k_z)|>|\Delta(k_z)|$ for all $k_z$ along the arc, then the surface arcs take the form of Fig.~\ref{fig:DiracCase}a.  If $|V_C(k_z)|>|\Delta(k_z)|$ for $k_z$ near each Dirac node, but $|V_C(k_z)|<|\Delta(k_z)|$ for some interval between the two Dirac nodes, then the surface arcs have a sharp corner at the Dirac nodes, but are reconstructed in the middle as shown in  Fig.~\ref{fig:DiracCase}b.  

On the other hand, if $|B|<|\Delta_0|$, the arcs are gapped out near the Dirac nodes.  If $|V_C(k_z)|<|\Delta(k_z)|$ for the entire range of $k_z$ between the bulk Dirac nodes, then there are no surface states at zero-energy Fig.~\ref{fig:DiracCase}d.  If, $|V_C(k_z)|<|\Delta(k_z)|$ only near the Dirac nodes, but $|V_C(k_z)|>|\Delta(k_z)|$ for some interval between the nodes, then a conventional surface state, decoupled from the bulk and with a smooth Fermi-surface can occur as shown in  Fig.~\ref{fig:DiracCase}c.

\subsection{Different Surface State Structure on Top and Bottom Surfaces}
In the main text, we have assumed that both the top and bottom surfaces of the Dirac SM exhibit the same surface-state structure.  In general, this need not be the case, as the two surfaces can have very different environments (e.g. the top surface could be exposed to vacuum while the bottom sits on a substrate).  

We have seen that the unconventional ``kinked" surface-states of Fig.~\ref{fig:DiracCase}a,b can exhibit quantized magnetic orbits involving the surface-arcs of both surfaces, connected through the bulk states.  What happens to these orbits when only one surface exhibits surface-states that are connected to the bulk?  This puzzle is resolved by noting that an electron can traverse the R-arc on the top surface, connect to the bulk chiral LL, propagate to the bottom surface, and reflect into the counter-propagating chiral LL via a virtual detour into the gapped surface states of the bottom surface.  The quantized energy levels stemming from this orbit and for various other combinations of surface-state structure can all be readily found by applying the semiclassical methods described in the main text.

%

%
%

\subsection{Magnetic-Breakdown in Dirac Surface States}
The magnetic field mixes the R and R' orbits, characterized byenergy scale $C\ell_B^{-2}$, can cause a state sliding along the R Fermi-arc to bypass the bulk nodal point and jump directly between the R' Fermi-arc and skip over the nodal region and detour into the bulk entirely.  The field scale at which such magnetic-breakdown type processes become important can be estimated as follows: the field `blurs' momentum by $\delta k\approx \ell_B^{-1}$, so the RR' mixing can only induces transitions within $k\lesssim \ell_B^{-1}$ of the bulk node.  An electron sliding across the arc spends time of order $\delta t \approx \frac{\delta k}{evB}\approx \ell_B$ in the transition region.  Magnetic-breakdown between R and R' is likely to occur if $C\ell_B^2\delta t\approx C\ell_B^{3}\approx  1$, defining a breakdown field scale $B_\text{break-down} \approx C^{2/3}/e$, This is a non-linear effect likely to become important only at large fields compared to the other relevant field-scales $B_\text{sat/mix}$.  

\begin{figure}[hhh]
\begin{center}
\includegraphics[width = 3in]{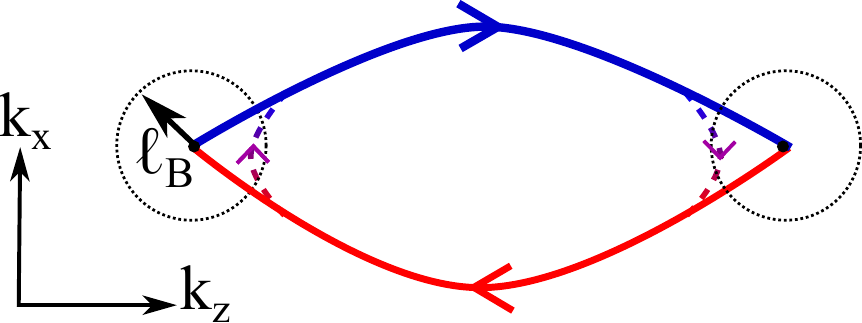}
\end{center}
\vspace{-.2in}
\caption{Magnetic break-down at high-fields can cause the state to skip over the bulk node (following the dashed-lines) creating oscillations purely within the surface arcs.
}
\label{fig:AppFig2}
\end{figure} 

For $B\gg B_\text{break-down}$, cyclotron motion can occur entirely in the surface arcs, with no detour into the bulk, corresponding to effectively cyclotron mass $m^\text{eff}_c\approx 2k_0/\pi v|\sin\phi|$ (compare to Eq.~\ref{eq:EffCyclotronMass}, there is no $\frac{1}{2\pi C}$ offset).

\section{\large Appendix E. Landau Level Structure of Dirac Semimetals}
We consider a Dirac semimetal in a tilted field which breaks the space group symmetry associated with the band degeneracy.  We show how, together with higher order (symmetry allowed) terms in the Dirac Hamiltonian, the tilted field mixes and gaps out the $n=0$ chiral Landau levels at each Dirac point.  Throughout this section, we parameterize the field in polar coordinates, with the polar axis, $\hat{z}$ chosen as the axis of symmetry that protects the Dirac nodes:
\begin{align} \v{B} = B\[\cos\theta\hat{z}+\sin\theta\(\cos\phi\hat{x}+\sin\phi\hat{y}\)\] \end{align}

\subsection{Low-Energy Hamiltonian for Na$_3$Bi  and Cd$_3$As$_2$}
While the space groups of Cd$_3$As$_2$ and Na$_3$Bi are different, they crucially both contain a symmetry operation of a screw-rotation around the $z$-axis. (For  Cd$_3$As$_2$ this is a four-fold $C_4$ rotation coupled to translation by $1/4$ a lattice vector; for Na$_3$Bi this is a six-fold $C_6$ rotation coupled to translation by $1/2$ a lattice vector.)  Thus for momentum points along the $\Gamma-Z$ line ($k_x=k_y=0$) the bands may be labelled as particular representations of the rotation symmetry. Any degeneracy between bands transforming as different representations is then protected by the crystal symmetry.  For both Cd$_3$As$_2$ and Na$_3$Bi it indeed happens that the conduction and valence bands transform as different representations ($J=1/2, J_z=\pm1/2$ and $J=3/2,J_z=\pm3/2$ Kramer's doublets) and do cross at $\vec{k}=\pm k_0 \hat{z}$, resulting in a pair of symmetry-protected Dirac points. 

The resulting low energy theories at these band crossings have an identical form for  Cd$_3$As$_2$ and Na$_3$Bi, differing only by particular values of parameters, and are given in Refs.~\cite{WangCd3As2,WangNa3Bi}. 
The low energy Hamiltonian for the $\vec{k}=+ k_0 \hat{z}$ Dirac point is $H=H_0+H'$, where
\begin{eqnarray}
& H_0 &=v \left[ k_x\mu^x\sigma^z - k_y\mu^y + k_z\mu^z \right] \\ 
& H' &= \alpha k_0 \mu^x \left[ (k_x^2 - k_y^2) \sigma^x + (k_x k_y+k_y k_x) \sigma^y \right]
 \end{eqnarray}
The Hamiltonian for the $- k_0 \hat{z}$ Dirac point is given by $k_0,k_z \rightarrow -k_0,-k_z$.  Here, $\mu$ matrices relate the two symmetry-distinguished bands ($\mu^3=\pm 1$ eigenstates correspond to $J=\frac{3}{2},\frac{1}{2}$ respectively), while $\sigma$ relate the Kramer's doublet degree of freedom for each band ($\sigma^3 = \pm $ corresponds to $\frac{J^z}{J}=\pm 1$).    For simplicity we have here taken the velocity $v$ at the Dirac point to be isotropic; any anisotropies can be absorbed into rescaling the coordinate system used to measure the magnetic field.  An important note on $H$ as written above is that the momentum $k_z$ is measured from the Dirac point; the displacement $K_z$ from $\Gamma$ is then $K_z = k_0 + k_z$. As we discuss below, this becomes an especially useful coordinate choice when we introduce a magnetic field, since we will take a magnetic field small enough so that $1/\ell_B << k_0$ and the two Dirac points don't mix.

\subsection{Choice of Landau level ladder operators}
Under minimal coupling $k\rightarrow k-q A$ to a magnetic field $B$, momentum operators gain a commutation relation $[k_a,k_b] = i q B_{a \times b}$. We consider a field tilted from the high symmetry $z$ axis by an angle $\theta$; The result is invariant under rotations in the $x,y$ plane.  Specifically, take the field to be 
\begin{equation}
\vec{B} = B\left[ \cos(\theta) \hat{z} + \sin(\theta)\hat{y} \right] .
\end{equation}
Then, one possible definition of ladder operators with the commutation relation $[a,a^\dagger]=1$ is the following:
\begin{equation}
a = \frac{\ell_B}{\sqrt{2}}\left[  k_x +i(\cos(\theta)k_y -\sin(\theta)k_z)  \right] .
\end{equation}
One can consider an alternative definition of an operator $\tilde{a}$ with $k_z$ replaced by $K_z = k_0 + k_z$, effectively shifted from the operator above by a constant. However as we show below, the Dirac point minimal Hamiltonian $H_0$ in a magnetic field can only be expressed simply in terms of the definition of $a$ above; its eigenstates are labelled by the number operator $\hat{n}=a^\dagger a$, but not by $\tilde{a}^\dagger \tilde{a}$. 

\subsection{Mixing of $n=0$ chiral modes}
We now show that for $\theta>0$, the zero energy chiral Landau levels mix and gap out.  This effect does not arise for $H_0$ alone, but rather requires higher order terms such as those appearing in $H'$.  We may thus work perturbatively in $H'$. 

First consider the spectrum for $H'=0$.  It is simplest to express the squares of the energies:
\begin{eqnarray}
& (H_0)^2 &=\frac{v^2}{\ell_B^2} \left[ (k_\parallel^2 \ell_B^2) + 
(2\hat{n}+1) + \cos(\theta) \mu^z \sigma^z - \sin(\theta) \mu^y \sigma^z
 \right] 
 \end{eqnarray}
Here $k_\parallel$ is the momentum quantum number parallel to the field; below we will be concerned with the gap at $k_\parallel = 0$. 
Landau level index $n$ (with the choice of ladder operator $a$ above) remains a good quantum number for $H_0$ at any field angle. 
The pair of counter-propagating $n=0$ modes remain degenerate.  Their two wavefunctions now mix the symmetry-protected representations,
\begin{eqnarray}
\left| \pm \right\rangle = \frac{1}{\sqrt{2\(1\mp \cos\theta\)}}\left| n=0 \right\rangle \otimes \left| \sigma^z=\pm1 \right\rangle \otimes
\left[  i(\cos(\theta)\mp 1)  \left| \mu^z=+1 \right\rangle + \sin(\theta) \left| \mu^z=-1 \right\rangle  \right] 
 \end{eqnarray}

Considering $H'$ as a perturbation, we find the effective Hamiltonian for the two $n=0$ states by evaluating the matrix elements of $H'$ between them. On the $n=0$ states, $H'$ projects to
\begin{eqnarray}
\left\langle n=0 |H' | n=0  \right\rangle = \frac{1}{2 \ell_B^2}\alpha k_0 \sin^2(\theta) \mu^x \sigma^x
 \end{eqnarray}
We find that the resulting gap $\Delta$ (double the magnitude of the $\left\langle + |H' | -  \right\rangle$ matrix element) is proportional to $B$, for any nonzero field angle:
\begin{eqnarray}
\Delta & = &  \alpha k_0 eB \sin^2\theta 
 \end{eqnarray}
Compared to the discussion of the main text, we identify $C = \alpha k_0$.


\begin{thebibliography}{10}
\bibitem{Endnote:SurfaceBulkTransition}
Alternative processes in which electrons transfer to non-chiral bulk LLs are not expected to play an important role in forming closed magnetic orbits since the non-chiral states 1) have finite energy gap $E_n(k_y)=\pm v_\perp\sqrt{2\frac{v_z}{v_\perp}\(eBn\)^2+ k_y^2}$, where $(|n|>0)$\cite{Endnote:SurfaceBulkTransition}, and 2) propagate  massively rather than ballistically.  This expectation is confirmed by the good agreement of semiclassical predictions with direct numerical simulations.
\bibitem{Endnote:LLTruncation}
This truncation breaks the canonical commutation relations of $a$ and $a^\dagger$ for states near the cutoff, $N_\text{max}$.  This can create spurious low-energy solutions, that have large weight on $N_\text{max}$ but very low energy.  These are removed by adding a regulator term that energetically penalizes $|N_\text{max}\>$.
\end{thebibliography}
\end{document}